\documentclass[12pt,twoside,english]{article}
\usepackage[T1]{fontenc}
\usepackage[latin1]{inputenc}
\usepackage{graphicx}
\usepackage[authoryear]{natbib}


\usepackage{babel}
\begin{document}

\title{Bath's law, correlations and magnitude distributions}

\author{{\normalsize{}B. F. Apostol }\\
{\normalsize{}Institute of Earth's Physics, Magurele-Bucharest MG-6,
}\\
{\normalsize{}POBox MG-35, Romania }\\
{\normalsize{}email: afelix@theory.nipne.ro}}

\date{{}}

\maketitle
\relax
\begin{abstract}
The empirical Bath's law is derived from the magnitude-difference
statistical distribution of earthquake pairs. It is shown that earthquake
correlations can be expressed by means of the magnitude-difference
distribution. We introduce a distinction between dynamical correlations,
which imply an \textquotedbl{}earthquake interaction\textquotedbl{},
and purely statistical correlations, generated by other, unknown,
causes. The pair distribution related to earthquake correlations is
presented. The single-event distribution of dynamically correlated
earthquakes is derived from the statistical fluctuations of the accumulation
time, by means of the geometric-growth model of energy accumulation
in the focal region. The derivation of the Gutenberg-Richter statistical
distributions in energy and magnitude is presented, as resulting from
this model. The dynamical correlations may account, at least partially,
for the roff-off effect in the Gutenberg-Richter distributions. It
is shown that the most suitable framework for understanding the origin
of the Bath's law is the extension of the statistical distributions
to earthquake pairs, where the difference in magnitude is allowed
to take negative values. The seismic activity which accompanies a
main shock, including both the aftershocks and the foreshocks, can
be viewed as fluctuations in magnitude. The extension of the magnitude
difference to negative values leads to a vanishing mean value of the
fluctuations and to the standard deviation as a measure of these fluctuations.
It is suggested that the standard deviation of the magnitude difference
is the average difference in magnitude between the main shock and
its largest aftershock (foreshock), thus providing an insight into
the nature and the origin of the Bath's law. It is shown that moderate-magnitude
doublets may be viewed as Bath partners. Deterministic time-magnitude
correlations of the accompanying seismic activity are also presented.
\end{abstract}
\relax

\section{Introduction }

Bath's law states that the average difference $\Delta M$ between
the magnitude of a main shock and the magnitude of its largest aftershock
is independent of the magnitude of the main shock (Bath, 1965; see
also Richter, 1958). The reference value of the average magnitude
difference is $\Delta M=1.2$. Deviations from this value have been
reported (see, for instance, Lombardi, 2002; Felzer et al, 2002; Console
et al, 2003), some being discussed by Bath, 1965. 

The Bath's law is an empirical statistical law. The earliest advance
in understanding its origin was made by Vere-Jones, 1969, who viewed
the main shock and its aftershocks as statistical events of the same
statistical ensemble, distributed in magnitude. The magnitude-difference
distribution introduced by Vere-Jones, 1969, may include correlations,
which are viewed sometimes as indicating the opinion that the main
shocks are statistically distinct from the aftershocks, or the foreshocks
(Utsu, 1969, Evison and Rhoades, 2001). The Bath's law enjoyed many
discussions and attempts of elucidation (Papazachos, 1974; Purcaru,
1974; Tsapanos, 1990; Kisslinger and Jones, 1991; Evison, 1999; Lavenda
and Cipollone, 2000; Lombardi, 2002; Helmstetter and Sornette, 2003).
The prevailing opinion ascribes the variations in $\Delta M$ to the
bias in selecting data and the insufficiency of the realizations of
the statistical ensemble. This standpoint was substantiated by means
of the binomial distribution (Console, 2003; Lombardi, 2002; Helmstetter
and Sornette, 2003). In order to account for the deviations of $\Delta M$
Helmstetter and Sornette, 2003, employed the ETAS (epidemic-type aftershock
sequence) model for the differences in the selection procedure of
the mainshocks and the aftershocks. According to this model the variations
in the number $\Delta M$ are related to the realizations of the statistical
ensemble and the values of the fitting parameters (see also Lombardi,
2002; Console, 2003).

The statistical hypothesis for the distributions of earthquake time
series is far-reaching. Usually, a statistical distribution is (quasi-)
independent of time (it is an equilibrium distribution), such that
the aftershocks, distributed in the \textquotedbl{}future\textquotedbl{},
should be identical with the foreshocks, distributed in the \textquotedbl{}past\textquotedbl{}.
However, it seems that there are differences between these two empirical
distributions (Utsu, 2002; Shearer, 2012). Moreover, in order the
statistical laws to be operational in practice, we need to have a
large statistical ensemble of accompanying (associated) earthquakes,
identified as aftershocks and foreshocks, which is a difficult issue
in empirical studies. In addition, the empirical realizations of the
statistical ensembles of earthquakes should be repetitive, a point
which raises problems of principle, considering the variations in
time and locations. In order to get meaningful results by applying
the statistical laws to physical ensembles, it is necessary to prepare,
in the same conditions, identical realizations of such ensembles.
This is possible in Statistical Physics for various physical systems
(like gases, liquids, solids,. etc). However, an ensemble of earthquakes,
collected from a given zone in a (long) period of time, is not reproducible,
because the conditions of its realization cannot be reproduced: the
distribution of the focal regions may change in the given zone, or
in the next period of time, including the change in physical (geological)
conditions. The application of the statistical method to earthquakes
exhibits limitations. 

We show in this paper that the appropriate tool of approaching the
accompanying seismic activity (foreshocks and aftershocks) of the
main shocks is the distribution function of the difference in magnitude.
The derivation of this distribution is made herein by means of the
conditional probabilities (the Bayes theorem), as well as by using
pair distributions. The earthquake correlations can be expressed by
means of the magnitude-difference distribution. These correlations
may be dynamical, or purely statistical. The dynamical correlations
arise from an \textquotedbl{}earthquake interaction\textquotedbl{},
while the purely statistical correlations originate in other, unknown,
causes. Pair distributions related to statistical correlations are
presented. The singe-event distribution of dynamically correlated
earthquakes is derived from the statistical fluctuations of the accumulation
time, by means of the geometric-growth model of energy accumulation
in the focal region. The dynamical correlations may account, at least
partially, for the roll-off effect in the Gutenberg-Richter statistical
distributions. The difference in magnitude is extended to negative
values, leading to a symmetric distribution for the foreshocks and
aftershocks, with a vanishing mean value for the magnitude difference.
This suggests to view the accompanying seismic activity as representing
fluctuations, and to take their standard deviation as a measure for
the Bath's average difference $\Delta M$ between the magnitude of
the main shock and its largest aftershock (foreshock). This way, the
Bath's law is deduced. It is shown that moderate-magnitude doublets
may be viewed as \textquotedbl{}Bath partners\textquotedbl{}. Deterministic
time-magnitude correlations of the associated seismic activity are
also presented.

\section{Gutenberg-Richter statistical distributions }

According to the geometric-growth model of energy accumulation in
a localized focal region (Apostol, 2006a,b), the accumulated energy
$E$ is related to the accumulation time $t$ by 
\begin{equation}
1+t/t_{0}=\left(1+E/E_{0}\right)^{r}\,\,\,,\label{1}
\end{equation}
 where $t_{0}$ and $E_{0}$ are time and energy thresholds and $r$
is a geometrical parameter which characterizes the focal region. This
parameter is related to the reciprocal of the number of effective
dimensions of the focal region and to the strain accumulation rate
(which, in general, is anisotropic). Very likely, the parameter $r$
varies in the range $1/3<r<1$. For a pointlike focal region with
a uniform accumulation rate $r=1/3$ (three dimensions), for a two-dimensional
uniform focal region $r=1/2$, while for a one-dimensional focal region
$r$ tends to unity. An average parameter $r$ may take any value
in this range. 

In equation (\ref{1}) the threshold parameters should be viewed as
very small, such that $t/t_{0},\,E/E_{0}\gg1$ and equation (\ref{1})
may be written as 
\begin{equation}
t/t_{0}\simeq\left(E/E_{0}\right)^{r}\,\,.\label{2}
\end{equation}
A uniform frequency of events $\sim t_{0}/t$ in time $t$ indicates
that the parameter $t_{0}$ is the reciprocal of the seismicity rate
$1/t_{0}$. It follows immediately the time distribution 
\begin{equation}
P(t)dt=\frac{1}{\left(t/t_{0}\right)^{2}}\frac{dt}{t_{0}}\label{3}
\end{equation}
 and, making use of equation (\ref{2}), the energy distribution 
\begin{equation}
P(E)dE=\frac{r}{\left(E/E_{0}\right)^{1+r}}\frac{dE}{E_{0}}\,\,.\label{4}
\end{equation}
 At this point we may use an exponential law $E/E_{0}=e^{bM}$, where
$M$ is the earthquake magnitude and $b=\frac{3}{2}\cdot\ln10=3.45$,
according to Kanamori, 1977, and Hanks and Kanamori, 1979 (see also,
Gutenberg and Richter, 1944, 1956); we get the (normalized) magnitude
distribution 
\begin{equation}
P(M)dM=\beta e^{-\beta M}dM\,\,\,,\label{5}
\end{equation}
 where $\beta=br$. In decimal logarithms, $P(M)=(1.5r)\cdot10^{-(1.5r)M}$,
where $0.5<1.5r<1.5$ (for $1/3<r<1$). Usually, the average value
$1.5r=1$ ($\beta=2.3$) is currently used as a reference value, corresponding
to $r=2/3$ (see, for instance, Stein and Wysesssion, 2003; Udias,
1999; Lay and Wallace, 1995; Frohlich and Davis, 1993). Since $t\gg t_{0}$,
$E\gg E_{0}$, the magnitude distribution given by equation (\ref{5})
is not adequate for very small magnitudes ($M\longrightarrow0$). 

It is worth noting that the magnitude distribution (equation (\ref{5}))
has the property $P(M_{1}+M_{2})\sim P(M_{1})P(M_{2})$, while the
time and energy distributions (equations (\ref{3}) and (\ref{4}))
have not this property. This is viewed sometimes as indicating that
the earthquakes would be correlated in occurrence time and in energy
(Corral, 2006). 

The magnitude distribution is particularly important because it can
be used to analyze the empirical distribution 
\begin{equation}
P(M)=\frac{\Delta N}{N_{0}\Delta M}=\frac{t_{0}\Delta N}{T\Delta M}\,\,\,,\label{6}
\end{equation}
of $\Delta N$ earthquakes with magnitude in the range $(M,\,M+\Delta M)$
out of a total number $N_{0}=T/t_{0}$ of earthquakes which occurred
in time $T$. We get 
\begin{equation}
\ln\left(\Delta N/T\right)=\ln\left(\frac{\beta\Delta M}{t_{0}}\right)-\beta M\,\,.\label{7}
\end{equation}

From the magnitude frequency $\Delta N/T$ (equation (\ref{6})) we
get the mean recurrence time

\begin{equation}
t_{r}=\frac{t_{0}}{\beta\Delta M}e^{\beta M}\label{8}
\end{equation}
for an earthquake with magnitude $M$ (\emph{i.e.} in the interval
$(M,\,M+\Delta M)$). This time should be compared with the accumulation
time $t_{a}=t_{0}e^{\beta M}$ for an earthquake with magnitude $M$,
given by equation (\ref{2}) and the exponential law $E/E_{0}=e^{bM}$.
These times are related by $t_{a}=(\beta\Delta M)t_{r}$, whence one
can see that $t_{a}<t_{r}$ (for $\beta\Delta M<1$), a relationship
which shows that the energy corresponding to a magnitude $M$ may
be lost by seismic events lower in magnitude, as expected. Moreover,
by the definition of the seismicity rate, an earthquake with magnitude
$M$ is equivalent with a total number $t_{a}/t_{0}=e^{\beta M}$
of earthquakes with zero magnitude (energy $E_{0}$) (Michael and
Jones, 1998; Felzer et al, 2002). We may call these earthquakes (defined
by the seismicity rate) \textquotedbl{}fundamental earthquakes\textquotedbl{}
corresponding to magnitude $M$. It is worth noting that the magnitude
distribution $\beta e^{-\beta M}$ implies an error of the order $\left(\sqrt{\overline{M^{2}}}-\overline{M}\right)/\overline{M}=\sqrt{2}-1$,
at least, \emph{i.e.}, $\Delta t_{r}/t_{r}\simeq0.41$. For a maximal
entropy with mean recurrence time $t_{r}$ we get easily a Poisson
distribution $(1/t_{r})e^{-t/t_{r}}$ for the recurrence time, which
has a (large) standard deviation $\sqrt{\overline{(t-t_{r})^{2}}}=t_{r}$. 

Similarly, from equation (\ref{5}) we get the excedence rate (the
so-called recurrence law), which gives the number $N_{ex}$ of earthquakes
with magnitude greater than $M$. The corresponding probability is
readily obtained from (\ref{5}) as $P_{ex}=e^{-\beta M}$, such that
the excedence rate reads
\begin{equation}
\ln N_{ex}=\ln N_{0}-\beta M\,\,.\label{9}
\end{equation}

The distributions given above may be called Gutenberg-Richter statistical
distributions (equations (\ref{7}) and (\ref{9}) and, implicitly,
(\ref{5})). They are currently used in statistical analysis of the
earthquakes. The parameter $\beta$ is derived by fitting these distributions
to data. For $1/3<r<1$ it varies in the range $1.15<\beta<3.45$
(for decimal logarithms $0.5<1.5r<1.5$). For instance, an analysis
of a large set of global earthquakes with $5.8<M<7.3$ ($\Delta M=0.1$)
indicates $\beta=1.38$ (and $1/t_{0}=10^{5.5}$ per year), corresponding
to $r=0.4$, a value which suggests an intermediate two/three-dimensional
focal mechanism (Bullen, 1963). For $r=1/3$, corresponding to a uniform
pointlike focal geometry, we get $\beta=1.15$. Equations (\ref{5}),
(\ref{7}) and (\ref{9}) have been fitted to a set of $1999$ earthquakes
with magnitude $M\geq3$ ($\Delta M=0.1$), which occurred in Vrancea
between $1974-2004$ ($31$ years) (Apostol 2006a,b). The mean values
of the fitting parameters are $-\ln t_{0}=9.68$ and $\beta=1.89$
($r=0.54$). A similar fit has been done for a set of $3640$ earthquakes
with magnitude $M\geq3$ which occurred in Vrancea during $1981-2018$
($38$ years). The fitting parameters for this set are $-\ln t_{0}=11.32$
and $\beta=2.26$ ($r=0.65$). The data for Vrancea have been taken
from the Romanian Earthquake Catalog, 2018. The parameter $\beta$
varies from region to region, depends on the nature of the focal mechanism
(parameter $r$ in $\beta=br$), the size and the time period of the
data set. The range of empirical values $0.5<1.5r<1.5$ coincides
with the theoretical range ($1/3<r<1$). 

The statistical analysis gives a generic image of a collective, global
earthquake focal region (a distribution of foci). Particularly interesting
is the parameter $r$, which is related to the reciprocal of the (average)
number of effective dimensions of the focal region and the rate of
energy accumulation. The value $r=0.54$ (Vrancea, period $1974-2004$)
indicates a (quasi-) two-dimensional geometry of the focal region
in Vrancea, while the more recent value $r=0.65$ for the same region
suggests an evolution of this (average) geometry towards one dimension.
At the same time, we note an increase of the seismicity rate $1/t_{0}$
in the recent period in Vrancea. The increase of the geometrical parameter
$r$ determines an increase of the parameter $\beta$, which dominates
the mean recurrence time. For instance, the accumulation time for
magnitude $M=7$ is increased from $t_{a}\simeq34.9$ years (period
$1974-2004)$ to at least $t_{a}\simeq59$ years. This large variability
indicates the great sensitivity of the statistical analysis to the
data set. In particular, for any fixed $M$ we may view the exponential
$Me^{-M\beta}$ as a distribution of the parameter $\beta$, which
indicates an error $\simeq0.41$ in determining this parameter. 

We note that inherent errors occur in statistical analysis. For instance,
an error is associated to the threshold magnitude (\emph{e.g.}, $M=3$),
because the large amount of data with small magnitude may affect the
fit. Also, it is difficult to include events with high magnitude in
a set with statistical significance, because such events are rare.
The size of the statistical set may affect the results. The fitting
values given above for Vrancea have an error of approximately $15\%$.
Such difficulties are carefully analyzed on various occasions (\emph{e.g.},
Felzer et al, 2002; Console et al, 2003; Lombardi, 2002; Helmstetter
and Sornette, 2003).

The statistical distributions given above may be employed to estimate
conditional probabilities, and to derive Omori laws for the associated
(accompanying) seismic activity. Also, the conditional probabilities
can be used for analyzing the next-earthquake distributions (inter-event
time distributions), (Apostol and Cune, 2020) which may offer information
for seismic hazard and risk estimation. We present here another example
of using these distributions, in analyzing the Bath's empirical law. 

\section{Bath\textquoteright s law }

In general, two or more earthquakes may appear as being associated
in time and space with, or without, a mutual interaction between their
focal regions. In both cases they form a foreshock-main shock-aftershock
sequence which exhibits correlations. The correlations which appear
as a consequence of an interaction imply an energy transfer (exchange)
between the focal regions (\emph{e.g}., a static stress). These correlations
may be called dynamical (or \textquotedbl{}causal\textquotedbl{})
correlations. Another type of correlations may appear without this
interaction, from unknown causes. For instance, an earthquake may
produce changes in the neighbourhood of its focal region (adjacent
regions), and these changes may influence the occurrence of another
earthquake. Similarly, an associated seismic activity may be triggered
by a \textquotedbl{}dynamic stress\textquotedbl{}, not a static one
(Felzer and Brodsky, 2006). \textquotedbl{}Unknown causes\textquotedbl{}
is used here in the sense that the model employed for describing these
earthquakes does not account for such causes. The correlations arising
from \textquotedbl{}unknown causes\textquotedbl{} may be called purely
statistical (or \textquotedbl{}acausal\textquotedbl{}) correlations.
It is worth noting that all the correlations discussed here have a
statistical character. 

Let us first discuss the dynamical correlations. If an amount $\delta E$
of energy (positive or negative) is provided to a focal region by
neighbouring focal regions, the accumulation time $t$ in the time-energy
accumulation law $t/t_{0}=(E/E_{0})^{r}$ (equation (\ref{2})) changes.
For a given energy we may assign this variation to the parameter $r$.
This may correspond to a change in the focal region subject to interaction
(as, for instance, one produced by a static stress). We are led to
examine the changes in the accumulation time $t$. We have shown above
that the ratio $t/t_{0}$ is the number $n_{0}=e^{\beta M}$ of \textquotedbl{}fundamental
earthquakes\textquotedbl{} corresponding to the magnitude $M$, where
$t_{0}$ is the cutoff time (time threshold; equations (\ref{2})
to (\ref{5})). We may write, approximately, this (large) number as
the sum 
\begin{equation}
n_{0}=\sum_{i=1}^{n_{0}}n_{i}\,\,,\,\,n_{i}=1\,\,.\label{10}
\end{equation}
In this equation we may view $n_{i}$ as statistical variables, with
mean value $\overline{n}_{i}=1$. Therefore, we consider the number
of fundamental earthquakes 
\begin{equation}
n=\sum_{i=1}^{n_{0}}n_{i}\,\,\,,\label{11}
\end{equation}
 with the mean value $\overline{n}=n_{0}$. The variables $n_{i}$
are viewed as independent statistical variables; their fluctuations
are due to the interaction of these fundamental earthquakes with other
fundamental earthquakes, corresponding to different magnitudes (exchange
of numbers of fundamental earthquakes). These interactions play the
role of an external \textquotedbl{}thermal bath\textquotedbl{} for
the fundamental earthquakes corresponding to a given magnitude $M$.
Therefore, the deviation $\Delta n=\sqrt{\overline{(\delta n)^{2}}}$
(where $\delta n=n-n_{0}$, $\delta n_{i}=n_{i}-\overline{n}_{i}$)
is the number $n_{c}$ of (dynamically) correlated fundamental earthquakes
corresponding to $M$. We get immediately 
\begin{equation}
\overline{(\delta n)^{2}}=\sum_{i,j=1}^{n_{0}}\overline{\delta n_{i}\delta n_{j}}=\sum_{i=1}^{n_{0}}\overline{(\delta n_{i})^{2}}=a^{2}n_{0}\,\,\,,\label{12}
\end{equation}
 where $a^{2}=\overline{(\delta n_{i})^{2}}$ (independent of $i$).
It follows 
\begin{equation}
n_{c}=a\sqrt{n_{0}}=ae^{\frac{1}{2}\beta M}\,\,.\label{13}
\end{equation}
 Obviously, this number is $t_{c}/at_{0}$, where $t_{c}$ is the
accumulation time for the correlated earthquakes with magnitude $M$
(in order to preserve the correspondence of the cutoff time and energy
parameters we need to re-define the cutoff time $t_{0}$ as $at_{0}$).
Now, we can repeat the derivation from equations (\ref{2}) to (\ref{5})
and get the magnitude probability of the correlated earthquakes 
\begin{equation}
P_{c}(M)=\frac{1}{2}\beta e^{-\frac{1}{2}\beta M}\,\,.\label{14}
\end{equation}
We note that in the energy-accumulation law given by equation (\ref{2})
the parameter $r$ changes to $r/2$ for dynamically correlated earthquakes
($t_{c}/at_{0}=(E/E_{0})^{r/2}$). 

The dynamically correlated earthquakes may be present in foreshock-main
shock-aftershock sequences. It is difficult to test empirically the
probability $P_{c}(M)$, because we cannot see any means to separate
the dynamical correlations from the purely statistical correlations.
We note that the dynamically correlated earthquakes are distinct from
the rest of the earthquakes (their distribution $\sim e^{-\frac{1}{2}\beta M}$
is different from the distribution $\sim e^{-\beta M}$). Usually,
in empirical studies we do not make this difference (see, for instance,
Kisslinger, 1996), but the error may be neglected, because the total
number of earthquakes $N_{0}$ is much larger than the total number
of dynamically correlated earthquakes $N_{c}$, according to equations
(\ref{5}), (\ref{6}) and (\ref{14}) ($N_{c}^{2}=(4\Delta N_{c}^{2}/\Delta N\Delta M)N_{0}$);
$N_{c}$ is, simply, proportional to the statistical error $\sqrt{N_{0}}$.
Like the total number of earthquakes, the dynamically correlated earthquakes
are concentrated in the region of small magnitudes, where the slope
of the function $\ln P_{c}(M)$ is changed from $-\beta$ to $-\frac{1}{2}\beta$.
Such a deviation is well known in empirical studies (the roll-off
effect; Pelletier, 2000; Bhattacharya et al, 2009), and is attributed
usually to an insufficient determination of the small-magnitude data.
Moreover, small-magnitude correlated earthquakes in foreshock-main
shock-aftershock sequences are associated with high-magnitude main
shocks, which have a large productivity of accompanying events. Therefore,
dynamically correlated earthquakes are expected in earthquake clusters
with high-magnitude main shocks. 

The Bath's law is expressed in terms of the difference in magnitude
between the main shock and its largest aftershock. Let us first consider
two earthquakes with magnitudes $M_{1,2}$, with the distribution
law $\sim e^{-\beta M}$, $M=M_{1,2}$, and seek the distribution
of the difference in magnitude $m=M_{1}-M_{2}$. In this law the magnitude
$M$ is positive, but for the difference $M_{1}-M_{2}$ we need to
extend this variable to negative values. Since $M_{1}=M_{1}-M_{2}+M_{2}$
and $M_{2}=M_{2}-M_{1}+M_{1}$, the law $\sim e^{-\frac{1}{2}\beta M}$
suggests a magnitude-difference distribution $\sim e^{-\beta(M_{1}-M_{2})}$
for $M_{1}>M_{2}$ and fixed $M_{2}$, and a distribution $\sim e^{-\beta(M_{2}-M_{1})}$
for $M_{2}>M_{1}$ and fixed $M_{1}$. These are conditional probabilities
(related to the Bayes theorem). In both cases, this distribution can
be written as $\sim e^{-\beta\mid m\mid}$, where $m=M_{1}-M_{2}$
(or $m=M_{2}-M_{1}$), $\mid m\mid<max(M_{1},M_{2})$, irrespective
of which $M_{1,2}$ is fixed. The condition $\mid m\mid<max(M_{1},M_{2})$
is essential for statistical correlations. We pass now to the Bath's
law. Let us assume a main shock, with magnitude $M_{s}$, and the
accompanying earthquakes (foreshocks and aftershocks), with magnitude
$M$. We define the foreshocks and aftershocks as those correlated
earthquakes with magnitude $M$ smaller than $M_{s}$. We refer the
magnitudes $M$ of the foreshocks and aftershocks to the magnitude
$M_{s}$ of the main shock, by observing the ordering of the partners
in each pair. This can be done by defining $m=M_{s}-M>0$ for foreshocks
and $m=M-M_{s}<0$ for aftershocks. According to the above discussion,
the distribution of the magnitude difference is $\sim e^{-\beta\mid m\mid}$,
$\mid m\mid<M_{s}$. Therefore, the total distribution is 
\begin{equation}
P(M_{s},m)=\beta^{2}e^{-\beta M_{s}}e^{-\beta\mid m\mid}\,\,,\,\,\mid m\mid<M_{s}\,\,,\,\,M_{s}>0\,\,.\label{15}
\end{equation}
 This is a pair distribution, for two events $M_{s}$ and $m$.

Let us apply first this law to dynamically-correlated earthquakes,
by replacing $\beta$ in equation (\ref{15}) by $\beta/2$. Since
these earthquake clusters are associated with high-magnitude main
shocks we may omit the condition $\mid m\mid<M_{s}$, and let $\mid m\mid$
go to infinity. In this case the statistical correlations are lost;
we are left only with the dynamical correlations. The distribution
given by equation (\ref{15}) becomes a distribution of two independent
events, identifed by $M_{s}$ and $m$; we may use only the magnitude
difference distribution 
\begin{equation}
p_{c}(m)=\frac{1}{4}\beta e^{-\frac{1}{2}\beta\mid m\mid}\,\,,\,\,-\infty<m<+\infty\,\,.\label{16}
\end{equation}
 This distribution has a vansihing mean value $\overline{m}$ ($\overline{m}=0$).
The next correction to this mean value, \emph{i.e.} the smallest deviation
of $m$, is the standard deviation 
\begin{equation}
\Delta m=\sqrt{\overline{m^{2}}}=\frac{2\sqrt{2}}{\beta}\,\,.\label{17}
\end{equation}
Therefore, we may conclude that the average difference in magnitude
between the main shock and its largest aftershock (or foreshock) is
given by the standard deviation $\Delta M=\Delta m=2\sqrt{2}/\beta$.
This is the Bath's law. The number $2\sqrt{2}/\beta$ does not depend
on the magnitude $M_{s}$ (but it depends on the parameter $\beta$,
corresponding to various realizations of the statistical ensemble).
It is worth noting that $\Delta m$ given by equation (\ref{17})
implies an averaging (of the squared magnitude differences). Making
use of the reference value $\beta=2.3$ we get $\Delta M=1.23$, which
is the Bath's reference value for the magnitude difference. In the
geometric-growth model the reference value $\beta=2.3$ corresponds
to the parameter $r=2/3$. We can check that higher-order moments
$\overline{m^{2n}}$, $n=2,3,...$ are larger than $\overline{m^{2}}$
(for any value of $\beta$ in the range $1.15<\beta<3.45$). 

The result $\Delta M=2\sqrt{2}/\beta$ could be tested empirically,
although, as it is well known, there exist difficulties. In empirical
studies the magnitude difference $\Delta M$ is variable, depending
on the fitting parameter $\beta$, which can be obtained from the
statistical analysis of the data. The results may tend to the value
$\Delta M=1.2$ by adjusting the cutoff magnitudes (Lombardi, 2002;
Console, 2003), or by choosing particular values of fitting parameters
(Helmstetter and Sornette, 2003); there are cases when the data exhibit
values close to $\Delta M=1.2$ (Felzer et al, 2002). It seems that
values closer to $\Delta M=1.2$ occur more frequently in the small
number of sequences which include high-magnitude main shocks.

If we extend the dynamical correlations to moderate-magnitude main
shocks, we need to keep the condition $\mid m\mid<M_{s}$, such that
the distribution is 
\begin{equation}
P_{c}(M_{s},m)=\frac{1}{4}\beta^{2}e^{-\frac{1}{2}\beta M_{s}}e^{-\frac{1}{2}\beta\mid m\mid}\,\,,\,\,\mid m\mid<M_{s}\,\,,\,\,M_{s}>0\,\,.\label{18}
\end{equation}
 The standard deviation is now $\Delta M=\Delta m=\sqrt{2}/\beta$,
which leads to $\Delta M\simeq0.61$ for the reference value $\beta=2.3$.
Such a variablility of $\Delta M$ can often be found in empirical
studies. For instance, from the analysis made by Lombardi, 2002, of
Southern California earthquakes 1990-2001 we may infere $\beta\simeq2$
and an average $\Delta M\simeq0.45$ (with large errors). From Console
et al, 2003, New Zealand catalog (1962-1999) and Preliminary Determination
of Epicentres catalog (1973-2001), we may infere $\beta\simeq2.5-2.3$
and an average $\Delta M=0.43-0.54$, respectively, while $\Delta M=\sqrt{2}/\beta$
gives $0.56-0.61$. In other cases, like the California-Nevada data
analyzed by Felzer et al, 2002, the parameters are $\beta=2.3$ and
$\Delta M\simeq1.2$, in agreement with $\Delta M=2\sqrt{2}/\beta$.
We note that $\Delta M=\sqrt{2}/\beta$ given here is an over-estimate,
because it extends, in fact, the dynamical correlations (equation
(\ref{14})) to small-magnitude main shocks.

Leaving aside the dynamical correlations we are left with purely statistical
correlations for clusters with moderate-magnitude main shocks. In
this case we use the distribution given by equation (\ref{15}), which
leads to $\Delta M=$$\Delta m=1/\sqrt{2}\beta$ and $\Delta M=0.31$
for the reference vale $\beta=2.3$. The Bath partner for such a small
value of the magnitude difference looks rather as a doublet (Poupinet
et al, 1984; Felzer et al, 2004).

For statistical correlations we can compute the correlation coefficient
(variance). The correlation coefficient $R=\overline{M_{s}M}/\Delta M_{s}\Delta M$
between the main shock and an accompanying event $M=M_{s}-\mid m\mid$
($\mid m\mid<M_{s}$) computed by using the distribution given in
equation (\ref{15}) is $R=2/\sqrt{5}$. For the correlation coefficient
beteen two accompanying events $M_{1}$ and $M_{2}$ we need the three-events
distribution (which includes $M_{1,2}$ and $M_{s}$). 

\section{Pair distribution }

In general, the correlations are visible in the pair (two-event, bivariate)
distributions. Such distributions are obtained as the mixed second-order
derivative of a generating function of two variables. We give here
the pair distribution derived from the geometric-growth model of energy
accumulation. We show that it coincides, practically, with the pair
distribution used above (equation (\ref{15})). Moreover, the pair
distribution derived here exhibits the dynamical correlations. The
single-event probability $P(t)$ given by equation (\ref{3}) is obtained
as the derivative $P(t)=-\partial F/\partial t$ of the frequency
function $F(t)=t_{0}/t$. Similarly, by using the change of variable
$t/t_{0}=e^{\beta M}$, we get the Gutenberg-Richter magnitude distribution
$P(M)=\beta e^{-\beta M}$ from $P(M)=-\partial F/\partial M$, where
$F(M)=e^{-\beta M}$ (equation (\ref{5})). 

Let us assume that two successive earthquakes may occur in time $t$,
one after time $t_{1}=t_{0}e^{\beta M_{1}}$, another after time $t_{2}=t_{0}e^{\beta M_{2}}$
from the occurrence of the former. Using the partition $t=t_{1}+t_{2}$
we get the distribution 
\begin{equation}
P(t_{1},t_{2})\sim\frac{\partial^{2}F}{\partial t_{1}\partial t_{2}}=\frac{2t_{0}}{(t_{1}+t_{2})^{3}}\,\,,\label{19}
\end{equation}
or, properly normalized, 
\begin{equation}
P(M_{1},M_{2})=4\beta^{2}\frac{e^{\beta(M_{1}+M_{2})}}{\left(e^{\beta M_{1}}+e^{\beta M_{2}}\right)^{3}}\,\,.\label{20}
\end{equation}
We can see that this distribution is different from $P(M_{1})P(M_{2})=\beta^{2}e^{-\beta(M_{1}+M_{2})}$,
which indicates that the two events $M_{1,2}$ are correlated. 

Let $M_{1}=M_{2}+m$ and $M_{1}>M_{2}$, $0<m<M_{1}$; equation (\ref{20})
becomes 
\begin{equation}
P(M_{1},M_{2})=4\beta^{2}\frac{e^{-\beta M_{1}}e^{-\beta m}}{\left(1+e^{-\beta m}\right)^{3}}\,\,;\label{21}
\end{equation}
 similarly, for $M_{2}>M_{1}$, $-M_{2}<m<0$ we get 
\begin{equation}
P(M_{1},M_{2})=4\beta^{2}\frac{e^{-\beta M_{2}}e^{\beta m}}{\left(1+e^{\beta m}\right)^{3}}\,\,.\label{22}
\end{equation}
 It follows that we may write 
\begin{equation}
P(M_{1},M_{2})=4\beta^{2}\frac{e^{-\beta max(M_{1},M_{2})}e^{-\beta\mid m\mid}}{\left(1+e^{-\beta\mid m\mid}\right)^{3}}\,\,\,,\label{23}
\end{equation}
which highlights the magnitude-difference distribution, with the constraint
$\mid m\mid<max(M_{1},M_{2})$. 

Let $M_{s}$ and $M$ be the magnitudes of the main shock and an acompanying
earthquake (foreshock or aftershock), respectively. We define the
ordered magnitude difference $m=M_{s}-M>0$ for foreshocks and $m=M-M_{s}<0$
for aftershocks, $\mid m\mid<M_{s}$. According to equation (\ref{23}),
the distribution of the pair consisting of the main shock and an acompanying
event is
\begin{equation}
P(M_{s},m)=4\beta^{2}e^{-\beta M_{s}}\frac{e^{-\beta\mid m\mid}}{\left(1+e^{-\beta\mid m\mid}\right)^{3}}\,\,,\,\,\mid m\mid<M_{s}.\label{24}
\end{equation}

The exponential $e^{-\beta\mid m\mid}$ falls off rapidly to zero
for increasing $m$, so we may neglect it in the denominator in equation
(\ref{24}). We are left with the pair distribution given by equation
(\ref{15}) (properly normalized). 

If we integrate equation (\ref{20}) with respect to $M_{2}$ (and
redefine $M_{1}=M$), we get the so-called marginal distribution 
\begin{equation}
P_{mg}(M)=\beta e^{-\beta M}\frac{2}{\left(1+e^{-\beta M}\right)^{2}}\,\,.\label{25}
\end{equation}
This distribution differs appreciably from the Gutenberg-Richter distribution
$\beta e^{-\beta M}$ for $\beta M$$\ll1$ and only slightly (by
an almost constant factor $\simeq2$) for moderate and large magnitudes.
The corresponding cummulative distribution for all magnitudes greater
than $M$ 
\begin{equation}
P_{mg}^{ex}(M)=e^{-\beta M}\frac{2}{1+e^{-\beta M}}\label{26}
\end{equation}
can be written as 
\begin{equation}
P_{mg}^{ex}(M)\simeq e^{-\beta M}\frac{1}{1-\frac{1}{2}\beta M}\simeq e^{-\frac{1}{2}\beta M}\label{27}
\end{equation}
 in the limit $M\longrightarrow0$, which indicates that the slope
of the excedence rate $\ln P_{mg}^{ex}(M)$ deviates from $-\beta$,
corresponding to the usual Gutenberg-Richter exponential distribution,
to $-\frac{1}{2}\beta$ (the roll-off effect). This deviation indicates
the presence of dynamical correlations governed by the distribution
law $P_{c}(M)=\frac{1}{2}\beta e^{-\frac{1}{2}\beta M}$ (equation
(\ref{14})).

The pair distribution given above can be written both for the earthquakes
governed by the Gutenberg-Richter distribution $\sim e^{-\beta M}$
and for the sub-set of dynamically-correlated earthquakes governed
by the distribution $\sim e^{-\frac{1}{2}\beta M}$. The procedure
of extracting dynamically-correlated earthquakes can be iterated,
passing from $\beta/2$ to $\beta/4$, etc; however, the number of
affected earthquakes tends rapidly to zero, and the procedure becomes
irrelevant. 

We may assume that energy $E$ is released by two successive earthquakes
with energies $E_{1,2}=E_{0}e^{bM_{1,2}}$, such that $E=E_{1}+E_{2}$.
The time corresponding to the energy $E$ is $t=t_{0}\left(E_{1}/E_{0}+E_{2}/E_{0}\right)^{r}$,
or $t=t_{0}\left(e^{bM_{1}}+e^{bM_{2}}\right)$. We cannot derive
a pair distribution from the second-order derivative of $t_{0}/t=\left(e^{bM_{1}}+e^{bM_{2}}\right)^{-1}$,
because the variation of the magnitudes $M_{1,2}$ implies the variation
of the energies $E_{1,2}$, and not of the times $t_{1,2}$. This
would contradict the geometric-growth model which assumes that the
probabilities are given by time derivatives. 

\section{Time-magnitude correlations}

Let us assume that the amount of energy $E$ accumulated in time $t$
is released by two successive earthquakes with energies $E_{1,2}$,
such as $E=E_{1}+E_{2}$. Since, according to equation (\ref{2}),
\begin{equation}
\begin{array}{c}
t/t_{0}=(E/E_{0})^{r}=\left(E_{1}/E_{0}+E_{2}/E_{0}\right)^{r}<\\
\\
<(E_{1}/E_{0})^{r}+(E_{2}/E_{0})^{r}=t_{1}/t_{0}+t_{2}/t_{0}\,\,\,,
\end{array}\label{28}
\end{equation}
 where $t_{1,2}$ are the accumulation times for the energies $E_{1,2}$,
we can see that the time corresponding to the pair energy is shorter
than the sum of the independent accumulation times of the members
of the pair, as expected for correlated earthquakes. This is another
type of correlations, different from dynamical or statistical correlations.
They are deterministic correlations, arising from the non-linearity
of the accumulation law given by equation (\ref{2}). The time interval
$\tau$ between the two successive earthquakes, 
\begin{equation}
\tau=t_{1}\left[\left(1+E_{2}/E_{1}\right)^{r}-1\right]\,\,\,,\label{29}
\end{equation}
 given by $t=t_{1}+\tau$, depends on the accumulation time $t_{1}$.
If we introduce the magnitudes $M_{1,2}$ in equation (\ref{29}),
we get 
\begin{equation}
\tau=t_{1}\left[\left(1+e^{-bm}\right)^{r}-1\right]\,\,\,,\label{30}
\end{equation}
 where $m=M_{1}-M_{2}$. We can see that this equation relates the
time $\tau$ to the magnitude difference. The same equation can be
applied to dynamically-correlated earthquakes, by replacing $r$ by
$r/2$. We get 
\begin{equation}
\tau=t_{1}\left[\left(1+e^{-bm}\right)^{r/2}-1\right]\,\,\,,\label{31}
\end{equation}
These correlations can be called time-magnitude correlations. 

We apply this equation to a main shock-aftershock sequence, where
$M_{1}$ is the magnitude of the main shock ($m>0$); similar results
are valid for the foreshock-main shock sequence. For the largest aftershock
(foreshock), where $m$ may be replaced by $\Delta m=2\sqrt{2}/\beta$,
we get 
\begin{equation}
\begin{array}{c}
\tau_{0}=t_{1}\left[\left(1+e^{-b\Delta m}\right)^{r/2}-1\right]\simeq\\
\\
\simeq\frac{1}{2}rt_{1}e^{-b\Delta m}=\frac{1}{2}rt_{1}e^{-2\sqrt{2}/r}
\end{array}\label{32}
\end{equation}
 (for $b\Delta m\gg1$). This is the occurrence time of the Bath partner,
measured from the occurrence of the main shock. The ratio $\tau_{0}/t_{1}$
varies between $3.5\times10^{-5}$ ($r=1/3$) and $3\times10^{-2}$
($r=1$); for $r=2/3$ we get $\tau_{0}/t_{1}=5\times10^{-3}$. 
\begin{figure}
\begin{centering}
\includegraphics[clip,scale=0.4]{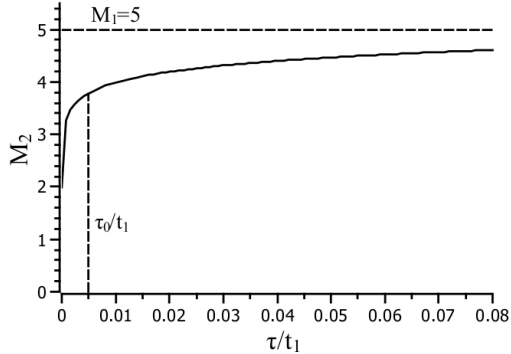}
\par\end{centering}
\caption{The magnitude $M_{2}$ of the accompanying seismic events \emph{vs}
the time $\tau$ elapsed from the main event with magnitude $M_{1}$
and accumulation time $t_{1}$ (equation (\ref{34}) for $M_{1}=5$,
$b=3.45$, $r=2/3$). The Bath partner $M_{2}\simeq3.8$ corresponds
to $\tau_{0}/t_{1}\simeq5\times10^{-3}$. Higher values of the magnitude
$M_{2}$ occur at much longer times, where the correlations are unlikely.
\label{fig:Fig.1}}
\end{figure}

It is worth noting, according to equation (\ref{28}), that a partner
close to the main shock in magnitude ($bm\ll1$) occurs after a lapse
of time 
\begin{equation}
\Delta t\simeq t_{1}\left(2^{r/2}-1\right)\,\,\,,\label{33}
\end{equation}
 which is much greater than $\tau_{0}$ ($\Delta t/t_{1}$ varies
between $0.12$ and $0.41$ for $1/3<r<1$). We can see that, even
if the pair probability $p(m)=(\beta/4)e^{-\frac{1}{2}\beta\mid m\mid}$
is greater for $m=0$, an earthquake close in magnitude to the main
shock occurs much later, where it may be difficult to view it as an
aftershock (and similarly for the foreshocks). Since $\left(1+e^{-bm}\right)^{r/2}$
(in equation (\ref{31})) is a decreasing function of $m$, we can
say, indeed, that the largest aftershock is farther in time with respect
to the main shock in comparison with aftershocks lower in magnitude.
The duration $\tau_{0}$ given by equations (\ref{32}) for the occurrence
of the largest aftershock may be taken as a measure of the extension
in time of the aftershock (and the foreshock) activity. It may serve
as a criterion for defining the accompaning seismic activity. 

Also, it is worth noting in this context that the (energy) probability
of two independent earthquakes with energies $E_{1,2}$ is $\sim P(E_{1})P(E_{2})$,
where $P(E_{1,2}$) is given by equation (\ref{4}) (similarly for
the time probability). On the other hand, if these two earthquakes
are correlated, then the conditional probability is obtained from
$P(E_{1}+E_{2})$ (as in equation (\ref{28})).

From equation (\ref{31}) we can get the distribution of the magnitudes
$M_{2}$ of the acompanying earthquakes with respect to the time $\tau$,
measured from the occurrence of the main shock with magnitude $M_{1}$,
either in the future (aftershocks) or in the past (foreshocks). Indeed,
we get
\begin{equation}
M_{2}=M_{1}+\frac{1}{b}\ln\left[\left(1+\tau/t_{1}\right)^{2/r}-1\right]\,\,\,,\label{34}
\end{equation}
 where $t_{1}$ ($=t_{0}e^{\frac{1}{2}\beta M_{1}}$) is the accumulation
time of the main shock; $M_{2}$ in equation (\ref{34}) is defined
for $\left(1+e^{-bM_{1}}\right)^{r/2}-1<\tau/t_{1}<2^{r/2}-1$ ($0<M_{2}<M_{1}$).
The function $M_{2}$ is plotted in Fig. \ref{fig:Fig.1} \emph{vs}
$\tau/t_{1}$ for $b=3.45$, $r=2/3$ ($\beta=2.3$) and $M_{1}=5$.
For $\tau/t_{1}$ very close to zero the magnitude $M_{2}$ is vanishing,
and for $\tau/t_{1}\longrightarrow2^{r/2}-1$ the magnitude $M_{2}$
tends to $M_{1}$; the Bath partner occurs at $\tau_{0}/t_{1}\simeq\frac{1}{2}re^{-2\sqrt{2}/r}\simeq5\times10^{-3}$
with the magnitude $M_{2}=3.8$. The function $M_{2}(\tau/t_{1})$
is a very steep function, for the whole (reasonable) range of parameters;
the whole accompanying seismic activity is, practically, concentrated
in the lapse of time $\tau_{0}$. On the scale $\tau/t_{1}$ the pair
probability of this activity is an abruptly increasing function of
$M_{2}$. If we use $M_{2}$ given by equation (\ref{34}) in the
distribution $\frac{1}{2}\beta e^{-\frac{1}{2}\beta M_{2}}$ for small
values of $\tau/t_{1}$, we get an Omori-type law, as expected (Apostol,
2006c). 

Finally, we note that the aftershock (foreshock) magnitude $M_{2}$
given by equation (\ref{34}) can be approximated by 
\begin{equation}
M_{2}\simeq(1-r/2)M_{1}+\frac{1}{b}\ln\left(\frac{2\tau}{rt_{0}}\right)\label{35}
\end{equation}
 for 
\begin{equation}
\frac{1}{2}re^{-(1-r/2)bM_{1}}<\tau/t_{0}<\left(2^{r/2}-1\right)e^{\frac{1}{2}\beta M_{1}}\,\,\,,\label{36}
\end{equation}
 where $t_{0}$ is the cutoff time. The lower bound $\frac{1}{2}rt_{0}e^{-(1-r/2)bM_{1}}$in
equation (\ref{36}) corresponds to a (very small) quiescence time
elapsed from the occurence of the main shock ($M_{2}=0$). This time
is much shorter than the cutoff time $t_{0}$, so, in fact, it is
irrelevant. The cumulative fraction of aftershocks (foreshocks) with
magnitude from zero to $M_{2}$ is $N_{cum}/N=1-e^{-\frac{1}{2}\beta M_{2}}$,
where $N$ is the total number of aftershocks (foreshocks). Making
use of equation (\ref{35}) we get the cumulative fraction for time
$\tau$ 
\begin{equation}
N_{cum}/N=1-\left(\frac{rt_{0}}{2\tau}\right)^{r/2}e^{-\frac{1}{2}(1-r/2)\beta M_{1}}\,\,.\label{37}
\end{equation}
This fraction is a rapidly increasing function of time, as it was
pointed out recently by Ogata and Tsuruoka, 2016. The cutoff time,
which is necessary in equation (\ref{37}), remains an empirical parameter.

\section{Concluding remarks }

In foreshock-main shock-aftershock sequences of associated (accompanying)
earthquakes we can discerne, in principle, two types of correlations.
One type, which we call dynamical (or causal) correlations, imply
an \textquotedbl{}interaction between earthquakes\textquotedbl{},
\emph{i.e.} an interaction between their focal regions (\emph{e.g.},
a static stress). Another type consists of purely statistical (or
acausal) correlations, though both types have a statistical character.
Dynamical correlations imply a variation (fluctuation) of the accumulation
time, which is estimated by means of the geometric-growth model of
energy accumulation in the focal region. This way, the number of dynamically
correlated earthquakes is derived, as a statistical standard deviation,
and the single-event distribution law for these earthquakes is deduced.
It is a Gutenberg-Richter-type exponential law with the parameter
$\beta$ changed into $\beta/2$. This change reflects a roll-off
effect in the Gutenberg-Richter statistical distribution, related,
mainly, to clusters with high-magnitude main shocks.

The correlations are discussed by means of the magnitude-difference
distribution for earthquake pairs, where the difference in magnitude
is extended to negative values. This distribution has a vanishing
mean value of the magnitude difference, such that the foreshock-aftershock
seismic activity appears as fluctuations in magnitude. The corresponding
standard deviation is the average diference in magnitudes between
the main shock and its greatest aftershock (foreshock). This difference
in magnitude is given by $\Delta M=2\sqrt{2}/\beta$ for dynamically
correlated earthquakes, which leads to $\Delta M=1.2$ for the reference
value $\beta=2.3$ ($1$ for decimal logarithms). This is the Bath's
law. If we allow for purely statistical correlations the difference
in magnitude is, approximately, $\Delta M=\sqrt{2}/\beta$. The difference
between these two formulae and the variability of the parameter $\beta$
may explain the variability in the results of the statistical analysis
of empirical data for $\Delta M$. We note that it is difficult to
differentiate empirically between dynamical and purely statistical
correlations. For earthquake clusters with moderate-magnitude main
shocks the purely statistical correlations dominate. They lead to
a Bath's law $\Delta M=1/\sqrt{2}\beta$, which gives $\Delta M=0.31$
for the reference value $\beta=2.3$, indicating rather doublets as
Bath's partners.

The roll-off effect in the Gutenberg-Richter distribution and the
presence of the dynamical correlations are also indicated by the pair
(two-event, bivariate) distribution, which is derived by means of
the geometric-growth model of energy accumulation. The same model
is used to derive the deterministic time-magnitude correlations occurring
in earthquake clusters. The time delay between the main shock and
its largest aftershock (foreshock) is estimated; it is suggested to
use this time interval as a criterion of estimating the temporal extension
of the aftershock (foreshock) activity. 

\textbf{Acknowledgements}

The author is indebted to the colleagues in the Institute of Earth\textquoteright s
Physics, Magurele, to members of the Laboratory of Theoretical Physics,
Magurele, for many enlightening discussions, and to the anonymous
reviewer for useful comments. This work was partially carried out
within the Program Nucleu 2016-2019, funded by Romanian Ministry of
Research and Innovation, Research Grant \#PN19-08-01-02/2019. Data
used for the fit (Vrancea region) have been extracted from the Romanian
Earthquake Catalog, 2018. Part of them (period 1974-2004) are available
through Apostol (2006a,b).

\textbf{REFERENCES}

Apostol, B. F. (2006a). A Model of Seismic Focus and Related Statistical
Distributions of Earthquakes. \emph{Phys. Lett.}, A357, 462-466 

Apostol, B. F. (2006b). A Model of Seismic Focus and Related Statistical
Distributions of Earthquakes. \emph{Roum. Reps. Phys.}, 58, 583-600 

Apostol, B. F. (2006c). Euler's transform and a generalized Omori
law. \emph{Phys. Lett.}, A351, 175-176 

Apostol, B. F. \& Cune, L. C. (2020). Short-term seismic activity
in Vrancea. Inter-event time distributions. \emph{Ann. Geophys.},
to appear

Bath, M. (1965). Lateral inhomogeneities of the upper mantle. \emph{Tectonophysics},
2, 483-514

Bhattacharya, P., Chakrabarti, C. K., Kamal \& Samanta, K. D. (2009).
Fractal models of earthquake dynamics. Schuster, H. G. ed. \emph{Reviews
of Nolinear Dynamics and Complexity} pp.107-150. NY: Wiley. 

Bullen, K. E. (1963). \emph{An Introduction to the Theory of Seismology}.
London: Cambridge University Press.

Console, R., Lombardi, A. M., Murru, M. \& Rhoades, D. (2003). Bath's
law and the self-similarity of earthquakes. \emph{J. Geophys. Res.},
108, 2128 10.1029/2001JB001651 

Corral, A. (2006). Dependence of earthquake recurrence times and independence
of magnitudes on seismicity history. \emph{Tectonophysics}, 424, 177-193 

Evison, F. (1999). On the existence of earthquake precursors. \emph{Ann.
Geofis}., 42, 763-770

Evison, F. \& and Rhoades, D. (2001). Model of long term seismogenesis.
\emph{Ann. Geophys.}, 44, 81-93

Felzer, K. R., Becker, T. W., Abercrombie, R. E., Ekstrom, G. \& Rice,
J. R. (2002). Triggering of the 1999 $M_{w}$ 7.1 Hector Mine earthquake
by aftershocks of the 1992 $M_{w}$ 7.3 Landers earthquake., \emph{J.
Geophys. Res.}, 107, 2190 10.1029/2001JB000911 

Felzer, K. R., Abercrombie, R. E. \& Ekstrom, G. (2004). A common
origin for aftershocks, foreshocks and multiplets. \emph{Bull. Seism.
Soc. Am.}, 94, 88-98

Felzer, K. R. \& Brodsky, E. E. (2006). Decay of aftershock density
with distance indicates triggerring by dynamic stress. \emph{Nature},
441,735-738

Frohich, C. \& Davis, S. D. (1993). Teleseismic $b$ values; or much
ado about $1.0$. \emph{J. Geophys. Res.}, 98, 631-644

Gutenberg, B. \& Richter, C. (1944). Frequency of earthquakes in California,
\emph{Bull. Seism. Soc. Am.}, 34, 185-188

Gutenberg, B. \& Richter, C. (1956). Magnitude and energy of earthquakes.
\emph{Annali di Geofisica}, 9, 1-15 ((2010)\emph{. Ann. Geophys.},
53, 7-12

Hanks, T. C. \& Kanamori, H. (1979). A moment magnitude scale. \emph{J.
Geophys. Res.}, 84, 2348-2350

Helmstetter, A. \& Sornette, D. (2003). Bath's law derived from the
Gutenberg-Richter law and from aftershock properties. \emph{Geophsy.
Res. Lett.}, 30, 2069 10.1029/2003GL018186 

Kanamori, H. (1977). The energy release in earthquakes. \emph{J. Geophys.
Res.}, 82, 2981-2987

Kisslinger, C. (1996). Aftershocks and fault-zone properties. \emph{Adv.
Geophys.}, 38 1-36

Kisslinger, C. \& Jones, L. M. (1991). Properties of aftershock sequences
in Southern California. \emph{J. Geophys. Res.}, 96, 11947-11958

Lavenda, B. H. \& Cipollone, E. (2000). Extreme value statistics and
thermodynamics of earthquakes: aftershock sequences. \emph{Ann. Geofis.},
43, 967-982

Lay, T. \& Wallace, T. C. (1995). \emph{Modern Global Seismology}.
San Diego, CA: Academic.

Lombardi, A. M. (2002). Probability interpretation of \textquotedbl{}Bath's
law. \emph{Ann. Geophys.}, 45, 455-472

Michael, A. J. \& Jones, L. M. (1998). Seismicity alert probability
at Parkfield, California, revisited. \emph{Bull. Seism. Soc. Am.},
88, 117-130

Ogata, Y. \& Tsuruoka, H. (2016). Statistical monitoring of aftershock
sequences: a case study of the 2015 $M_{w}$7.8 Gorkha, Nepal, earthquake.
\emph{Earth, Planets and Space}, 68:44, 10.1186/s40623-016-0410-8

Papazachos, P. C. (1974). On certain aftershock and foreshock parameters
in the area of Greece. \emph{Ann. Geofis.}, 24, 497-515

Pelletier, J. D. (2000). Spring-block models of seismicity: review
and analysis of a structurally heterogeneous model coupled to the
viscous asthenosphere. Rundle, J. B., Turcote, D. L. \& Klein, W.
eds. \emph{Geocomplexity and the Physics of Earthquakes}. vol. 120.
NY: Am. Geophys. Union. 

Poupinet, G., Elsworth, W. L. \& Frechet, J. (1984). Monitoring velocity
variations in the crust using earthquake doublets: an application
to the Calaveras fault, California. \emph{J. Geophys. Res}., 89, 5719-5731

Purcaru, G. (1974). On the statistical interpretation of the Bath's
law and some relations in aftershock statistics. \emph{Geol. Inst.
Tech. Ec. Stud. Geophys. Prospect.} (Bucharest), 10, 35-84

Richter, C. F. (1958). \emph{Elementary Seismology} (p.69). San Francisco,
CA: Freeman.

\emph{Romanian Earthquake Catalogue} (ROMPLUS Catalog). (2018). National
Institute for Earth Physics, Romania.

Shearer, P. M. (2012). Self-similar earthquake triggering, Bath's
law, and foreshock/aftershock magnitudes: simulations, theory, and
results for Southerm California. \emph{J. Geophys. Res.}, 117, B06310,
10.1029/2011JB008957

Stein, S. \& Wysession, M. (2003). \emph{An Introduction to Seismology,
Earthquakes, and Earth Structure}. NY: Blackwell.

Tsapanos, T. M. (1990). Spatial distribution of the difference between
magnitudes of the main shock and the largest aftershock in the circum-Pacific
belt. \emph{Bull. Seism. Soc. Am.}, 80, 1180-1189

Udias, A. (1999). \emph{Principles of Seismology}. NY: Cambridge University
Press.

Utsu, T. (1969). Aftershocks and earthquake statistics (I,II): Source
parameters which characterize an aftershock sequence and their interrelations.\emph{
J. Fac. Sci. Hokkaido Univ.}, Ser. VII, 2, 129-195, 196-266

Utsu, T. (2002). Statistical features of seismicity. \emph{International
Geophysics}, 81, Part A, 719-732

Vere-Jones, D. (1969). A note on the statistical interpretation of
Bath's law. \emph{Bull. Seismol. Soc. Amer.}, 59, 1535-1541
\end{document}